\documentclass[pdftex,twocolumn,epjc3]{svjour3}          

\RequirePackage[T1]{fontenc}

\smartqed  

\RequirePackage{graphicx}
\RequirePackage{mathptmx}      
\RequirePackage{flushend}
\RequirePackage[numbers,sort&compress]{natbib}
\RequirePackage[colorlinks,citecolor=blue,urlcolor=blue,linkcolor=blue]{hyperref}

\usepackage{soul,xcolor}
\usepackage[latin9]{inputenc}
\usepackage{amsmath}
\usepackage{amssymb}
\usepackage{amsfonts}
\usepackage{lscape}
\usepackage{hyperref}
\usepackage{url}
\usepackage{color}
\usepackage{bm}
\usepackage{tabularx}
\usepackage{mathtools}
\usepackage{ulem}
\newcommand{\be}{\begin{equation}}
\newcommand{\ee}{\end{equation}}
\newcommand{\ba}{\begin{eqnarray}}
\newcommand{\ea}{\end{eqnarray}}

\renewcommand{\[}{\begin{equation}}
\renewcommand{\]}{\end{equation}}
\def\be{\begin{equation}}
\def\ee{\end{equation}}
\def\bea{\begin{eqnarray}}
\def\eea{\end{eqnarray}}
\def\eqi{\begin{equation}}
\def\eqf{\end{equation}}
\def\eqia{\begin{eqnarray}}
\def\eqfa{\end{eqnarray}}

\definecolor{darkgreen}{rgb}{0,0.6,0}
\definecolor{violet}{rgb}{1.5,0,1.5}

\journalname{Eur. Phys. J. C}

\begin{document}

\setstcolor{red}

\title{Can the angular scale of cosmic homogeneity be used as a cosmological test?}


\author{Xiaoyun Shao\thanksref{e1,addr1}
        \and
        Rodrigo S. Gon\c{c}alves\thanksref{e2,addr1,addr2} 
        \and
        Carlos A. P. Bengaly\thanksref{e3,addr1}
        \and
        Uendert Andrade\thanksref{e4,addr3}
        \and
        Gabriela C. Carvalho\thanksref{e5,addr4}
        \and
        Jailson Alcaniz\thanksref{e6,addr1}
}

\thankstext{e1}{e-mail: xiaoyun48@on.br}
\thankstext{e2}{e-mail: rsousa@on.br}
\thankstext{e3}{e-mail: carlosbengaly@on.br}
\thankstext{e4}{e-mail: uendsa@umich.edu}
\thankstext{e5}{e-mail: gabriela.coutinho@fat.uerj.br}
\thankstext{e6}{e-mail: alcaniz@on.br}

\institute{Observat\'orio Nacional, 20921-400, Rio de Janeiro, RJ, Brazil\label{addr1}
          \and
          Departamento de F\'isica, Universidade Federal Rural do Rio de Janeiro, 23897-000, Serop\'edica, RJ,  Brazil\label{addr2}
          \and
          Department of Physics, University of Michigan, 450 Church St, Ann Arbor, MI 48109-1040, USA\label{addr3}
          \and
          Faculdade de Tecnologia, Universidade do Estado do Rio de Janeiro,  27537-000, Resende, RJ, Brazil\label{addr4}
}

\date{Received: date / Accepted: date}

\maketitle

\begin{abstract}
In standard cosmology, the cosmic homogeneity scale is the transition scale above which the patterns arising from non-uniformities -- such as groups and clusters of galaxies, voids, and filaments -- become indistinguishable from a random distribution of sources. Recently, different groups have investigated the feasibility of using such a scale as a cosmological test and arrived at different conclusions. In this paper, we complement and extend these studies by exploring the evolution of the spatial ($R_{\rm H}$) and angular ($\theta_{\rm H}$) homogeneity scales with redshift, assuming a spatially flat, $\Lambda$-Cold Dark Matter universe and linear cosmological perturbation theory. We confirm previous results concerning the non-monotonicity of $R_{\rm H}$ with the matter density parameter $\Omega_{\rm m0}$ but also show that it exhibits a monotonical behavior with the Hubble constant $H_{0}$ within a large redshift interval. More importantly, we find that, for $z \gtrsim 0.6$, $\theta_{\rm H}$ presents a monotonical behavior with $\Omega_{\rm m0}$, as well as for $H_0$ the entire redshift interval analyzed. We find also that the angular homogeneity scale is  sensitive to $H_{0}$, especially at higher redshifts. Using the currently available $\theta_{\rm H}$ measurements, we  illustrate the constraints on the $\Omega_{\rm m0}$ - $H_{0}$ plane through a MCMC analysis and show the feasibility of using the angular homogeneity scale as a new, model-independent way to constrain cosmological parameters.
\end{abstract}

\section{Introduction}
\label{sec:intro}

The flat $\Lambda$-Cold Dark Matter ($\Lambda$CDM) model, where $\Lambda$ represents the cosmological constant, has been established as the standard cosmological model (SCM) for over two decades. Although this model is able to provide a successful explanation to a variety of cosmological  observations~\cite{aghanim2021planck,Brout:2022vxf,eBOSS:2020yzd,DES:2021wwk,ACT:2023kun,Li:2023tui}, it is plagued with several theoretical caveats, e.g. coincidence and fine-tuning problems~\cite{Weinberg:2000yb}, in addition to recent observational issues, such as the $\sim5\sigma$ tension between early and late-time measurements of the Hubble constant (see e.g. \cite{DiValentino:2021izs} and references therein).

In light of these issues, it is paramount to test the foundations of the SCM, since any significant departure from their assumptions would require a complete reformulation of such paradigm. One of these hypothesis is the Cosmological Principle (CP), which states that the Universe looks statistically homogeneous and isotropic at sufficiently large scales, although lumpy and inhomogeneous at smaller scales due to the presence of cosmic structures like clusters, voids and filaments. Therefore, we can describe cosmological distances and ages by means of the Friedmann-Lema\^itre-Robertson-Walker (FLRW) metric from such scale onwards, as in the case of the SCM. Nonetheless, it should be stressed that we can only directly test the assumption of statistical isotropy, not the homogeneity one. This happens because we are only able to observe cosmic sources on the intersection between our past light cone and the time-constant spatial hypersurfaces where they are located - hence, we can only perform consistency tests of this latter assumption. See~\cite{Clarkson:2010uz, Maartens:2011yx, Clarkson:2012bg} for a broad discussion on this topic.

Given the advent of large redshift surveys, numerous tests were carried out in order to measure the cosmic scale where the Universe appears consistent with the CP, i.e., the scale where the actual three-dimensional (3D) distribution of cosmic sources becomes statistically indistinguishable from a random one. Most of these tests could identify and measure such a characteristic scale, namely the {cosmic homogeneity scale}, $R_{\rm H}$, at around $70-150 \; \mathrm{Mpc}$ using a variety of galaxy and quasar catalogues from these redshift surveys~\cite{Hogg:2004vw, Sarkar:2009iga, Scrimgeour:2012wt, Pandey:2013xz, Pandey:2015xea, Sarkar:2016fir, Laurent:2016eqo, Ntelis:2017nrj, Goncalves:2018sxa, Goncalves:2020erb, Kim:2021osl}, albeit some works claimed otherwise~\cite{Labini:2009ke, Labini:2011dv, Park:2016xfp}, besides that these measurements could be biased by the survey window function~\cite{Heinesen:2020wai}.

Still, these measurements are only able to provide a { consistency test} of the SCM and the CP, as we need to convert the source redshifts into distances - and hence an assumption of a fiducial cosmological model must be made. This issue can be circumvented by using the {{angular homogeneity scale}}, $\theta_{\rm H}$, i.e., the two-dimensional (2D) measure of the homogeneity scale, given only in terms of the source position in the sky. This analysis was originally proposed by~\cite{Alonso:2013boa}, being subsequently performed by~\cite{Alonso:2014xca, Goncalves:2017dzs, Andrade:2022imy} using different redshift survey catalogues.

Recently, it was proposed that the 3D homogeneity scale $R_{\rm H}$ could be used as a sort of standard ruler, in a similar fashion to the sound horizon scale in baryonic acoustic oscillations measurements. Hence, it would be feasible to test cosmological models and constrain their parameters using $R_{\rm H}$ measurements~\cite{Ntelis:2018ctq, Ntelis:2019rhj}. However, a subsequent paper showed that such scale cannot provide a valid standard ruler, due to its non-monotonic relation with the matter density parameter~\cite{Nesseris:2019mlr}. Still, it is worth mentioning that none of these works examined the feasibility of measurements of the angular (2D) homogeneity scale as a possible cosmological test. 

The goal of this paper is to study the behavior of angular homogeneity scale $\theta_{\rm H}$ with redshift and investigate whether it can provide a one-to-one relation with respect to different values of cosmological parameters and the clustering bias, conversely from its three-dimensional counterpart, $R_{\rm H}$. 

The structure of our paper is outlined as follows: Section~\ref{Sec:meth_theo_fram} describes the methodology of the homogeneity scale within the standard cosmological model paradigm, the theoretical framework and the estimators, as well as the criteria commonly adopted to measure the homogeneity scale with observational data. Section~\ref{sec:numerical_results} is dedicated to the numerical outcomes of our analysis of the homogeneity scale, as described in the two previous sections. Finally, section~\ref{sec:conclusions} provides a summary of our study and the conclusions that we have drawn.


\section{Methodology}
\label{Sec:meth_theo_fram}

This section provides a brief description of our method. For a complete review on the quantities we cite~\cite{Hogg:2004vw, Sarkar:2009iga, Scrimgeour:2012wt, Pandey:2013xz, Pandey:2015xea, Sarkar:2016fir, Laurent:2016eqo, Ntelis:2017nrj, Goncalves:2018sxa, Goncalves:2020erb, Kim:2021osl,Labini:2009ke, Labini:2011dv, Park:2016xfp,Heinesen:2020wai,Alonso:2013boa,Alonso:2014xca, Goncalves:2017dzs, Andrade:2022imy,Ntelis:2018ctq, Ntelis:2019rhj,Nesseris:2019mlr,gonccalves2018cosmic} and references therein.

\subsection{3D homogeneity scale, $R_{\rm H}$}
\label{3d2.1}

The 3D statistical homogeneity scale is obtained through the scaled count-in-spheres, defined as the number $N(<r)$ of objects in a galaxy catalog inside a spherical shell of radius $r$, compared to the number $N_{\rm random}(<r)$ in a similar shell of
a mock uniformly random catalog, defined as

\begin{equation}
\mathcal{N}(<r)=\frac{N(<r)}{N_{\rm random}(<r)} \,,
\label{e11}
\end{equation}
where geometric effects and completeness of the survey are taken into consideration using $N_{\mathrm{random}}(<r)$. Assuming that the Universe is statistically isotropic and homogeneous, the number of objects in a properly constructed and homogeneous mock catalogue will scale as $N_{\mathrm{random}}(<r) \propto r^{3}$, -- implicitly assuming that the observational catalogue is properly complete, homogeneous and clean of possible impurities e.g. stellar contamination -- whereas the number of objects in the real catalogue will generally scale as $N(<r) \propto r^{D_{2}}$, where $D_{2}$ is the fractal index, which takes the value $D_{2}=3$ for a purely homogeneous catalogue.

By using the logarithmic derivative of the scaled count-in-spheres, which is provided by Eq.~\ref{e11}, one can compute the fractal index $D_{2}$ according to
\begin{equation}
D_{2}(r)=3+\frac{d \ln \mathcal{N}(<r)}{d \ln r}.
\label{e22}
\end{equation}

The value of the 3D fractal dimension should asymptotically approach the number of environmental dimensions at scales where the underlying distribution becomes statistically homogeneous. In other words, we should expect that
\begin{equation}
D_{2}(r) \rightarrow 3 \quad \mathrm{for} \quad r \rightarrow R_{\rm H}.
\end{equation}
Usually, the scale of homogeneity is defined as the scale $R_{\rm H}$ where the fractal index $D_{2}$ reaches $1\%$ of the expected value~\cite{Scrimgeour:2012wt}, so that
\begin{equation}
D_{2}\left(R_{\rm H}\right)=2.97 \;.
\label{e33}
\end{equation}
Detailed calculations of $R_{\rm H}$ are shown in
~\cite{Nesseris:2019mlr}. Although the definition of $1\%$ for the homogeneity scales is arbitrary it does not depend on the survey and can be used to test cosmological models and compare different survey measurements as long as the same definitions are used in all cases~\cite{Scrimgeour:2012wt}. See also ~\cite{Laurent:2016eqo} for an analysis with a different threshold.

Additionally,~\cite{Ntelis:2018ctq}~presented the idea that the 3D homogeneity scale $R_{\rm H}$, as defined by Eq.~\ref{e33}, could be used as a standard ruler like the Baryon Acoustic Oscillations (BAO). This is an appealing concept in theory, since $R_{\rm H}$ is readily ascertained from galaxy catalogues. However, $R_{\rm H}$ is model-dependent as a standard ruler to constrain cosmological parameters, because it needs to adopt a fiducial cosmological model to convert the redshifts of galaxies into cosmological distances. Thus, in what follows, we will introduce the angular homogeneity scale $\theta_{\rm H}$, which observationally does not require a cosmological model because we only need the celestial coordinates of the galaxies to obtain it.

\subsection{2D homogeneity scale, $\theta_{\rm H}$}
$ $

In order to obtain an observational measurement of the angular homogeneity scale using the 2D fractal dimension $D_{2}$, we must first define the integral correlation $\mathcal{N}$. This quantity can be expressed in terms of the count-in-spheres compared to the average number of neighboring galaxies around a data point in a synthetic random catalogue. The former consists on the number of neighboring galaxies in a 2D-sphere of radius $\theta$ around a given data point, $N(<\theta)$, with $\theta$ denoting the angular separation on the sky between the pairs, and the latter is defined by $N_{\textrm{random}}(<\theta)= 2 \pi \bar{\rho}(1-\cos \theta)$, with the same angular scale $\theta$.

Hence, we can write the scaled counts-in-spheres as
\begin{equation}
\label{eq:defNr}
\mathcal{N}(<\theta)\equiv \frac{N(<\theta)}{N_{\textrm{random}}(<\theta)}, 
\end{equation}
where $N_{\textrm{random}}(<\theta)$ is used to account for geometric effects and completeness of the survey, as in the 3D case~\cite{Scrimgeour:2012wt}.

Assuming that the Universe is statistically isotropic and homogeneous, the number of objects scales as $N_{\textrm{random}}(<\theta)\propto \theta^2$, for small angles $\theta$, while in general the number of objects in a real catalogue will scale as $N(<\theta) \propto \theta^{D_{2}}$. Thus, accordingly to the 3D case, the determination of the 2D fractal index $D_{2}$ involves the utilization of the logarithmic derivative of the scaled counts-in-spheres, i.e.,
\begin{equation}
\label{eq:defD2}
D_{2}(\theta)=2+\frac{d \ln \mathcal{N}(<\theta)}{d \ln \theta} \,. 
\end{equation}

Similarly, the value of the 2D fractal dimension should asymptotically approach the number of environmental dimensions at scales where the underlying distribution becomes statistically homogeneous. In other words, we should expect that \cite{gonccalves2018cosmic}
\begin{equation}
D_{2}(\theta) \rightarrow 2 \quad \mathrm{for} \quad \theta \rightarrow \theta_H, \;\; \forall \theta \leq 20^{\circ} \,.
\label{eq:d2th}
\end{equation}

We also define that the angular homogeneity characteristic scale is reached where $D_{2}$ approaches 1\% of its expected value, following~\cite{Scrimgeour:2012wt}. This implies that $\theta_H$ represents the angular scale where
\begin{equation}
\label{e1.98}
D_{2}(\theta_H) = 1.98 \,.
\end{equation}
We can obtain a theoretical prediction of $\theta_H$ that satisfies the condition shown above for a given set of cosmological parameters, as we shall present in the following section.


\subsection{Theoretical Framework} 
$ $

In this and next sections, we show the relevant theoretical quantities by assuming the 2D analysis, as it is the main core of the novel results. The equivalence to the 3D analysis is briefly mentioned at the end of the next section.

The conditional probability of finding a galaxy in the element of solid angle $\delta \Omega$ at distance $\theta$ from a randomly chosen galaxy in the ensemble is \cite{peebles1980large}
\begin{equation}
\delta P=\bar{\rho} \delta \Omega[1+\omega(\theta)],
\label{pro2}
\end{equation}
where $\bar{\rho}$ and $\omega(\theta)$ are the mean density of galaxies and the two-point angular correlation
function (2PACF), respectively. 
Then, the expected number of neighbors within distance $\theta$ of a galaxy is
\begin{equation}
P(<\theta) = \bar{\rho} \int d \Omega[1+\omega(\theta)]. 
\end{equation}
Hence, by taking that the probability of a distribution of points is related by its number of points, we can rewrite the scaled counts-in-spheres, Eq.~\ref{eq:defNr}, as~\cite{Andrade:2022imy}
\begin{equation}
\label{eq:defNr2}
\mathcal{N}(<\theta) = 1+\frac{1}{1-\cos \theta} \int_{0}^{\theta} \omega\left(\theta^{\prime}\right) \sin \theta^{\prime} d \theta^{\prime},
\end{equation}
whereas the integral correlation can be associated to the 2PACF via the combination of Eq.~\ref{eq:defD2} and Eq.~\ref{eq:defNr2}, which reads
\begin{equation}
\label{eq:d2xi}
D_{2}(\theta)=2+\frac{d \ln}{d \ln \theta}\left[1+\frac{1}{1-\cos \theta} \int_{0}^{\theta} \omega\left(\theta^{\prime}\right) \sin \theta^{\prime} d \theta^{\prime}\right].
\end{equation}
From the theoretical perspective the 2PACF can be expressed as~\cite{crocce2011modelling}
\begin{equation}
\omega(\theta)=\int d z_{1} f\left(z_{1}\right) \int d z_{2} f\left(z_{2}\right) \xi\left(r\left(z_{1}\right), r\left(z_{2}\right), \theta, \bar{z}\right)\label{eq:d2w1}
\end{equation}
where $\xi$ is the matter three-dimensional two-point correlation function, and we assume $f(z) \equiv b(z) \phi(z)$, with  $b(z)$ being  the bias parameter and $\phi(z)$ the radial selection, which  is the probability to include a galaxy in a given redshift bin. It is worthful to mention that, even though the 3D correlation function ($\xi$) appears in the previous equation, the $D_2$ is obtained for the 2D case, as $\xi(r)$ leads to $\omega(\theta)$.

Also, the comoving distance to redshift $z$ in Eq.~\ref{eq:d2w1} is given by 
\begin{equation}
r(z)=\int_{0}^{z} \frac{c}{H(z')} dz',
\label{eq:rz}
\end{equation}
where  
\begin{equation}
H(z)=H_0\sqrt{\Omega_{\rm m0}(1+z)^{3} + (1-\Omega_{\rm m0})}\;.
\label{eq:Hz}
\end{equation}

In this paper, we will compute the 2PACF by assuming Limber approximation and integrating as in~\cite{simon2007accurate, shao2022probing}:
\begin{equation}
\omega(\theta) \approx \int_{0}^{\infty} d r_{1} \int_{0}^{\infty} d r_{2} P_{1}\left(r_{1}\right) P_{2}\left(r_{2}\right) \xi(R),\label{eq:d2w2}
\end{equation}
where
\begin{equation}
R \equiv \sqrt{r_{1}^{2}+r_{2}^{2}-2 r_{1} r_{2} \cos \theta}
\label{eq:e28}
\end{equation}
and $P_{1}(r_{1})$, $P_{2}(r_{2})$ correspond to the comoving radial distance distribution of the sample. 

\begin{figure}[!t]
	\centering
	\includegraphics[width=0.3\textwidth]{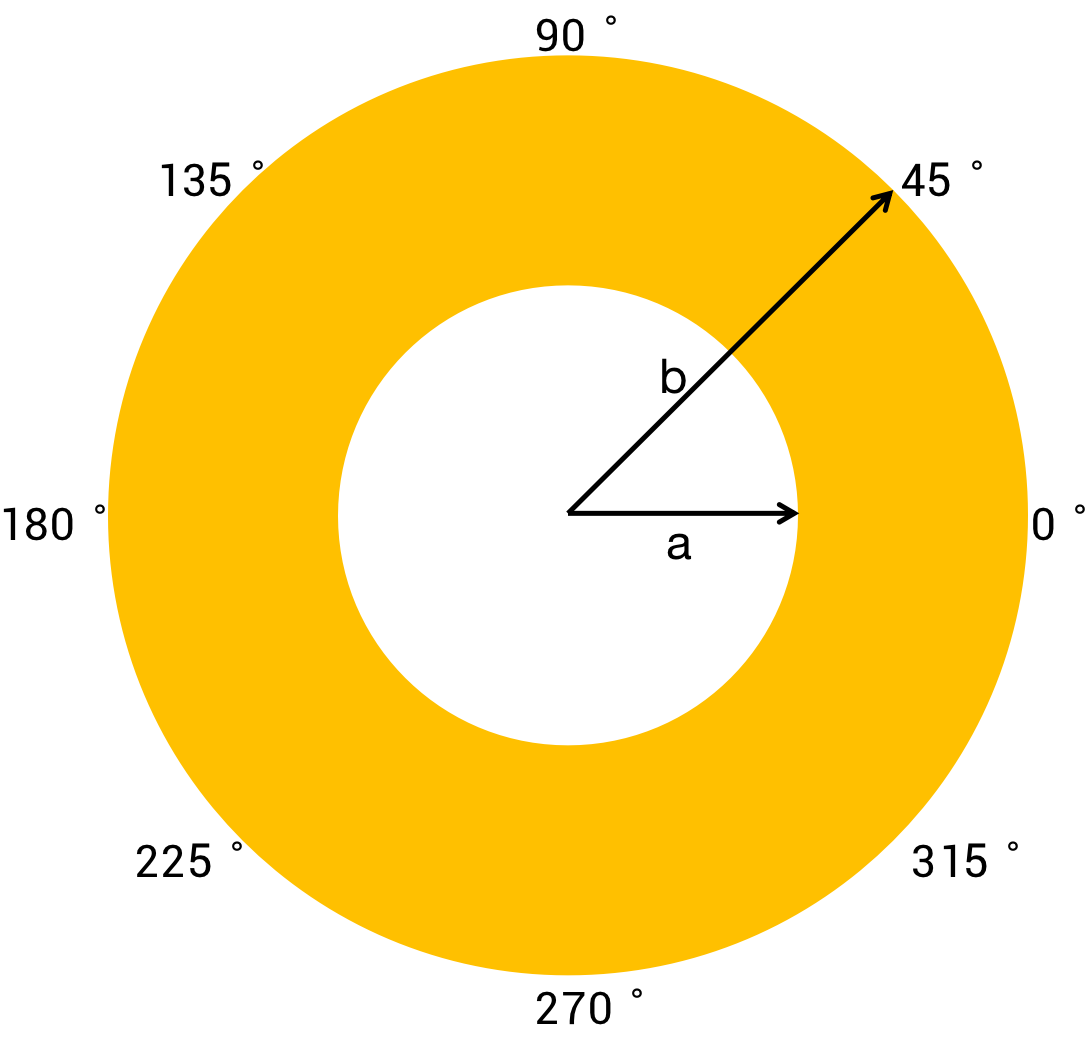}
	\caption{An illustration of the two-dimension shell where the 2PACF is computed. Angular: RA(degrees), Radial: redshift. a and b are the inner and outer radius of the spherical shell.}
	\label{fig:shell}
\end{figure}

Since we only consider top-hat window functions and narrow redshift bins, these quantities can be expressed by
\begin{equation}
P_{1}=P_{2}=P(r)\approx\left\{\begin{array}{l}0 \quad 0 \leq r<a \\ \frac{1}{b-a} \quad a \leq r<b \\ 0 \quad r \geq b\end{array}\right.
\end{equation}
where $a$ and $b$ are the inner and outer radius, respectively, of the spherical shell where the 2PACF is to be computed, as shown in Figure~\ref{fig:shell}.

The 3D two-point correlation function $\xi(r)$ can be expressed by
\begin{equation}
\xi(r)=\frac{1}{2\pi^2}\int_0^\infty j_0(k r) k^2 P(k) dk,\label{eq:corfunc}
\end{equation}
where the matter power spectrum, denoted as $P(k)$, is related to the spherical Bessel function $j_0(x)=\sin{(x)}/x$. We use the CAMB code to calculate the theoretical prediction of the matter power spectrum $P(k)$~\cite{lewis2000efficient}. In this work, our focus is solely on large cosmological scales, thus restricting our analysis to the linear segment of the spectrum. Also by doing this restriction we ensure the threshold of $1\%$ for $D_2$, since a larger threshold should take into account non-linear modelling of $P(k)$~\cite{Ntelis:2019rhj}.

Therefore this section showed we can obtain a theoretical prediction of $\theta_H$ for a given set of cosmological parameters. Now, in order to make a phenomenological study and to constrain the cosmological parameters we must obtain an estimation of a theoretical quantity, and fit those theoretical parameters with the observational quantities, as follows.


\subsection{Observational Estimator}
$ $

The key quantity to define an estimator for the angular homogeneity scale consists on the 2D fractal dimension $D_{2}(\theta)$ \cite{Alonso:2013boa,Goncalves:2017dzs,Andrade:2022imy}. We can estimate this quantity directly from data and random catalogues, by substituting in Eq.~\ref{eq:defNr2} the $\omega(\theta)$ for the Landy-Szalay 2PACF, $\hat{\omega}_{ls}(\theta)$~\cite{landy1993bias}, defined as
\begin{equation}\label{eq:w_ls}
\widehat{\omega}_{ls}(\theta) = \frac{DD(\theta) -2DR(\theta) + RR(\theta)}{RR(\theta)},
\end{equation}
and $DD(\theta)$, $DR(\theta)$, and $RR(\theta)$ are numbers of pairs as a function of separation $\theta$, normalized by the total number of pairs, in data-data, data-random and random-random catalogues, respectively. 

So the estimator for the scaled count-in-spheres in 2D is given by
\begin{equation}
\widehat{\mathcal{N}}(<\theta)=1+\frac{1}{1-\cos \theta} \int_{0}^{\theta} \widehat{\omega}_{l s}\left(\theta^{\prime}\right) \sin \theta^{\prime} d \theta^{\prime},
\label{est}
\end{equation}
then the 2D fractal dimension is obtained similarly to Eq.~\ref{eq:d2xi}.

\vspace{1cm}

The 2D analysis can be generalized to 3D by taking, from the beginning, the conditional probability of finding a galaxy, in the element of volume $\delta V$, at a distance $r$ from a randomly chosen galaxy in the ensemble as $\delta P=\bar{\rho} \delta V[1+\xi(\theta)]$. Thus $\mathcal{N}(<r)$ and $D_2(r)$ can be calculated from the two point galaxy correlation function $\xi(r)$~\cite{Nesseris:2019mlr}. With all this quantities we can perform our statistical analyses, where we use a fiducial cosmology with all cosmological parameters fixed at the best-fit Planck 18 flat $\Lambda$CDM model~\cite{aghanim2021planck}, except the matter density parameter ($\Omega_{\rm m0}$) and the reduced Hubble constant $h$ ($h \equiv H_0/(100 \; \mathrm{km \, s}^{-1} \,\mathrm{Mpc}^{-1})$) which we set as free parameters in our analysis.

\section{Numerical Results}
\label{sec:numerical_results}

This Section is dedicated to bring up a work of the literature, confirm its specific results, as well as extend it for a more general scenario and show novel results. With effect, the authors of \cite{Nesseris:2019mlr} showed that the 3D homogeneity scale (i.e., the value of $R_{\rm H}$ where $D_{2} (R_{\rm H}) = 2.97$), could not be a feasible standard ruler, as it does not show a one-to-one relation with a variation specifically on the matter density parameter $\Omega_{\rm m0}$ in the $\Lambda$CDM model. Moreover, the feasibility of a one-to-one relation was not investigated for other cosmological parameters neither for measurements of the angular homogeneity scale $\theta_H$. Thus, in what follows, we confirm their results, extend the analysis to various redshifts, clustering bias values, and cosmological parameter values, within the standard model framework. After we emphasize the angular homogeneity scale case with new and important results to the discussion.

\subsection{Spatial homogeneity scale}
\label{sec:rh_results}

\begin{figure*}[!ht]
	\centering
	\includegraphics[width=0.49\textwidth]{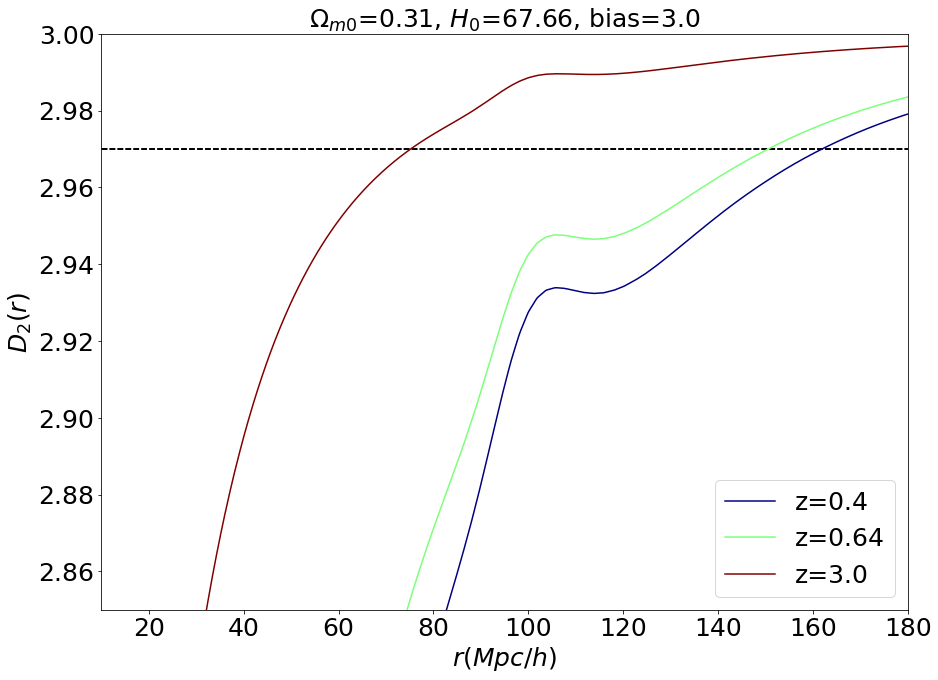}
    \includegraphics[width=0.49\textwidth]{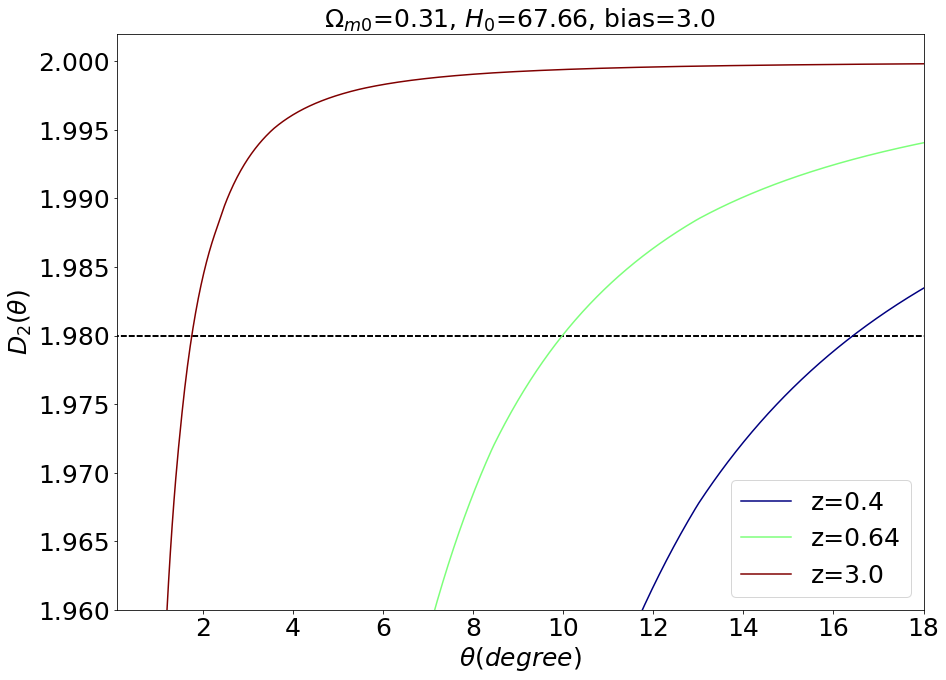}
	\caption{The 3D fractal dimension $D_{2}(r)$ (left) and the 2D fractal dimension $D_{2}(\theta)$ (right) for selected values of $z$ assuming $\Omega_{\rm m0}=0.31$ and $H_{0}=67.66$. The dashed line corresponds to $1\%$ deviation from homogeneity for both cases, i.e., $D_{2}=2.97$ and $D_{2}=1.98$, respectively.}
	\label{fig:3d0.4ksi}
\end{figure*}

\begin{figure*}[!ht]
	\centering
	\includegraphics[width=0.33\textwidth]{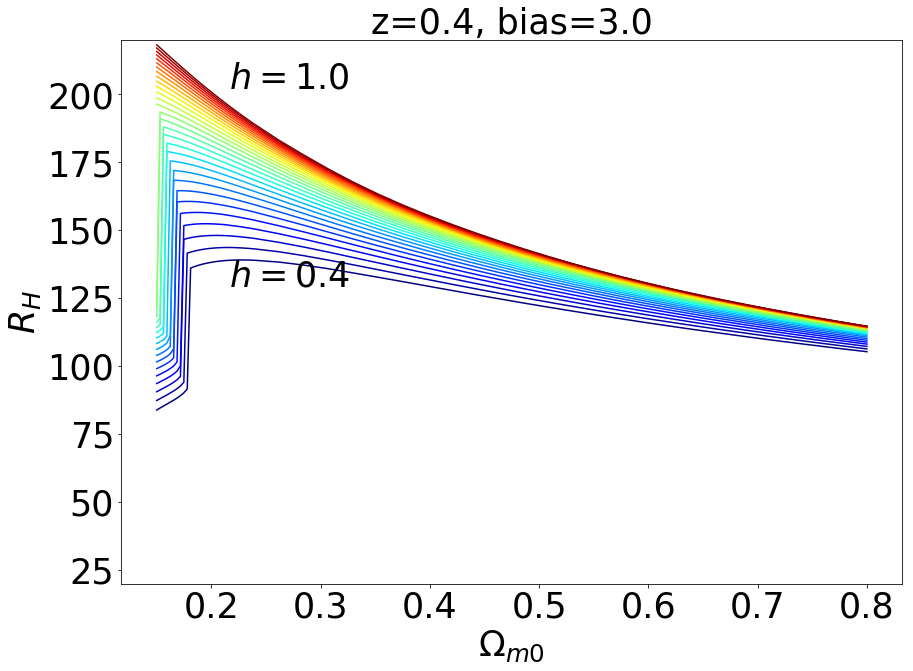}
	\includegraphics[width=0.33\textwidth]{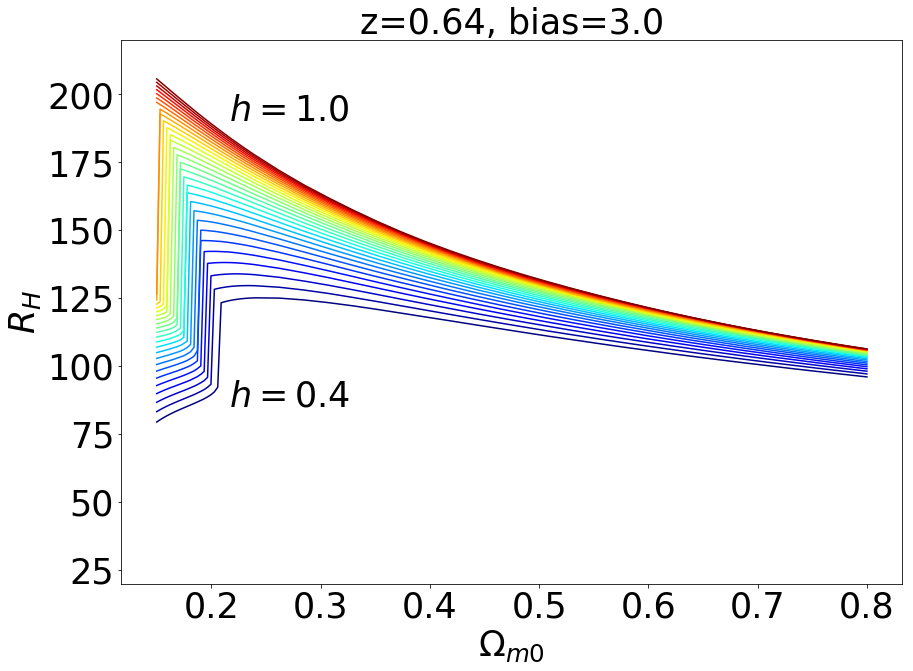}
 	\includegraphics[width=0.33\textwidth]{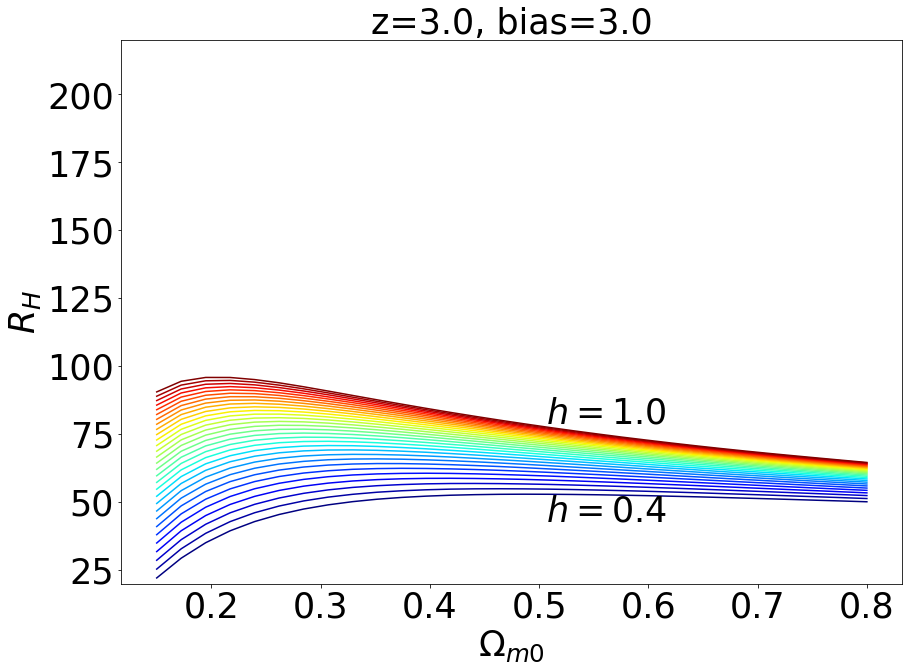}
  	\caption{The 3D homogeneity scale $R_{\rm H}$ as a function of the present matter density $\Omega_{\rm m0}$ for $h \in [0.4,1.0]$ at $z=0.4$, $0.64$, $3.0$. The 3D homogeneity scale $R_{\rm H}$ doesn't exhibit a  monotonical behavior for all values of $\Omega_{\rm m0}$ in the range considered. The sudden change in $R_{\rm H}$ around $\Omega_{\rm m0}$=0.2 is due to the BAO feature results in a non-smooth behavior of homogeneity scale \cite{ntelis2019cosmological}.}
	\label{fig:3d0.4hom_om}
\end{figure*}

\begin{figure*}[!ht]
	\centering
	\includegraphics[width=0.33\textwidth]{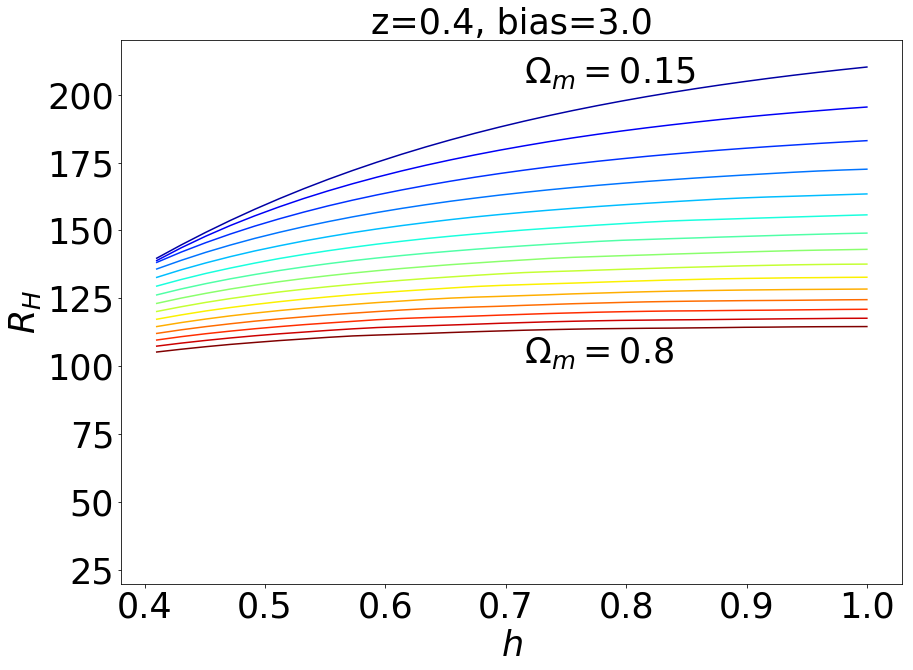}
	\includegraphics[width=0.33\textwidth]{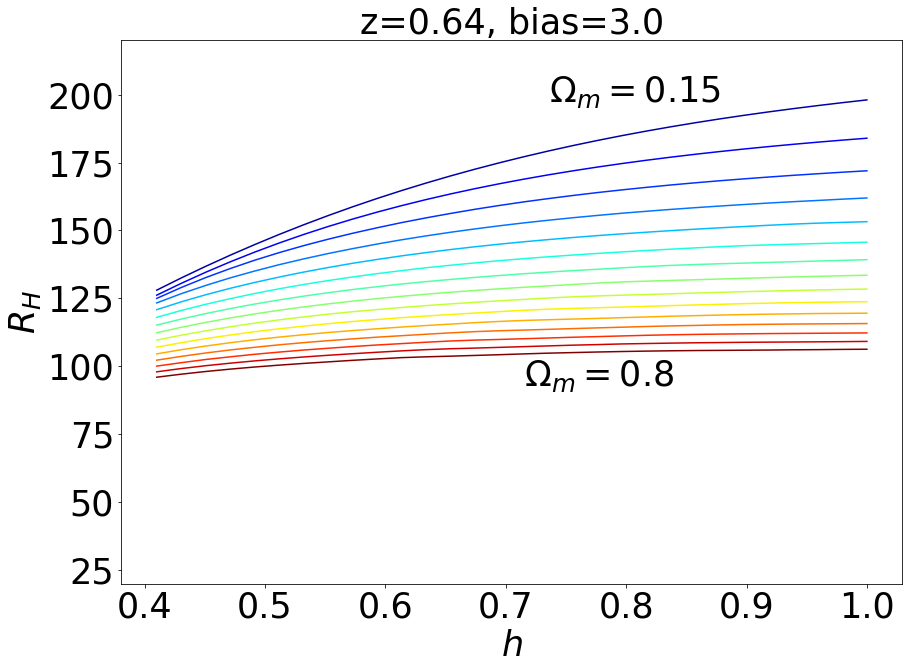}
 	\includegraphics[width=0.33\textwidth]{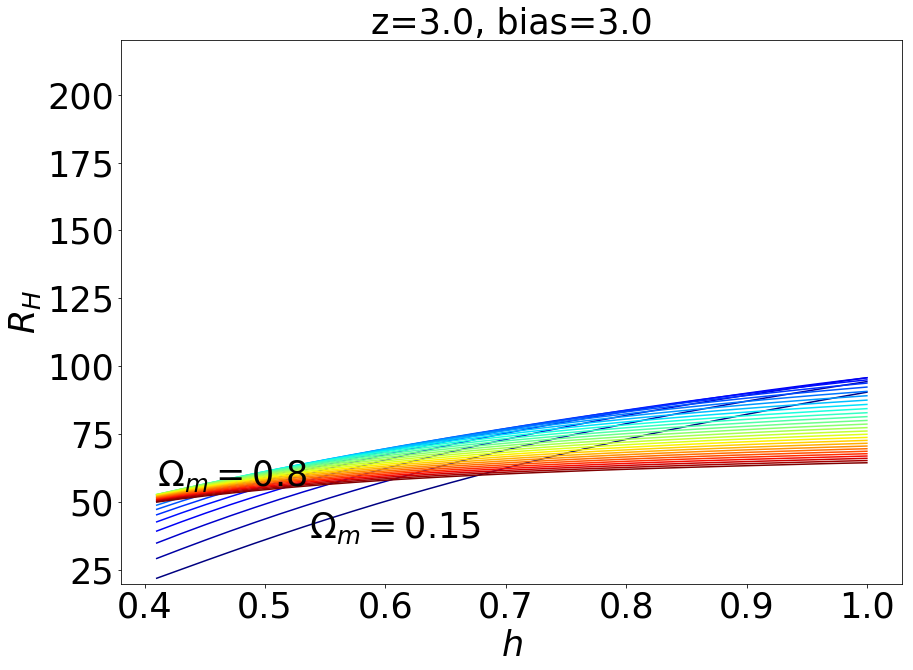}
  	\caption{The 3D homogeneity scale $R_{\rm H}$ as a function of the Hubble Constant $h$ for $\Omega_{\rm m0} \in [0.15,0.8]$ at $z=0.4$, $0.64$, $3.0$. The 3D homogeneity scale $R_{\rm H}$ exhibits a  monotonically increasing behavior for all values of $h$ in the range considered.}
	\label{fig:3d0.4hom}
\end{figure*}

First, we compute the theoretical predictions of the three-dimensional (spatial) homogeneity scales $R_{\rm H}$ from Eq.~\eqref{e22} and Eq.~\eqref{e33}, by the same token of~\cite{Nesseris:2019mlr}. Figure~\ref{fig:3d0.4ksi} (left panel) shows the 3D fractal dimension $D_{2}(r)$ for three redshift values, i.e., $z=0.4$, $0.64$, and $3.0$, assuming $\Omega_{\rm m0}=0.31$ and $h=0.677$, as consistent with the Planck 2018 best-fit~\cite{aghanim2021planck}. The dashed line corresponds to the $D_{2}=2.97$ condition, where the cosmic three-dimensional scale $R$ can be assigned as $R_{\rm H}$, as discussed in section~\ref{3d2.1}.

After, we check the monotonicity between $R_{\rm H}$ and $\Omega_{\rm m0}$ (Fig.~\ref{fig:3d0.4hom_om}), and we find that it indeed exhibits a non-monotonic behaviour for $\Omega_{\rm m0}=[0.15,0.8]$ values, which agrees with previous results showed in~\cite{Nesseris:2019mlr}. Note that we change the value of $\Omega_{\rm m0}$ by varying the cold dark matter density $\Omega_{\rm c0}$ and keeping the baryon density constant at $\Omega_{\rm b0} = 0.048$, based on the best-fit from Planck 18 \cite{aghanim2021planck}.
Therefore, our results confirm that the scale of cosmic homogeneity cannot be used as a cosmological probe to measure the cosmological parameter $\Omega_{\rm m0}$.

In order to extend the previous result, we investigate the monotonicity of the dimensionless Hubble Constant $h$. Fig.~\ref{fig:3d0.4hom} displays the 3D homogeneity scale $R_{\rm H}$ as a function of the $h$, considering the same redshift values as before, and for a range of $h=[0.4,1.0]$. All plots assume a clustering bias value of $b=3.0$. The results show that $R_{\rm H}$ exhibits a one-to-one, monotonically increasing function for all values of $h$ assumed. Therefore, our results show that the scale of cosmic homogeneity can be used as a cosmological probe to measure the cosmological parameter $h$.

As for additional remarks, we show that the 3D homogeneity scale $R_{\rm H}$ exhibits significant change with respect to reasonable values of $\Omega_{m0}$, i.e., from $0.21$ to $0.41$. We find a $\sim11.5\%$, $\sim11.7\%$, $\sim2.1\%$ relative change at redshift $z=0.4$, $0.64$, and $3.0$, respectively, for $\Omega_{m0}$ within this range. Similar behavior is found for the the 3D homogeneity scale $R_{\rm H}$ with respect to different $h$ values, as we obtain a $\sim3.7\%$, $\sim4.3\%$, and $\sim9.8\%$ difference at redshift $z=0.4$, $0.64$, and $3.0$, respectively, for $h$ within the range $[0.6,0.8]$.

\subsection{Angular homogeneity scale}
\label{sec:thetah_results}
$ $

\begin{figure*}[!ht]
	\centering
	\includegraphics[width=0.33\textwidth]{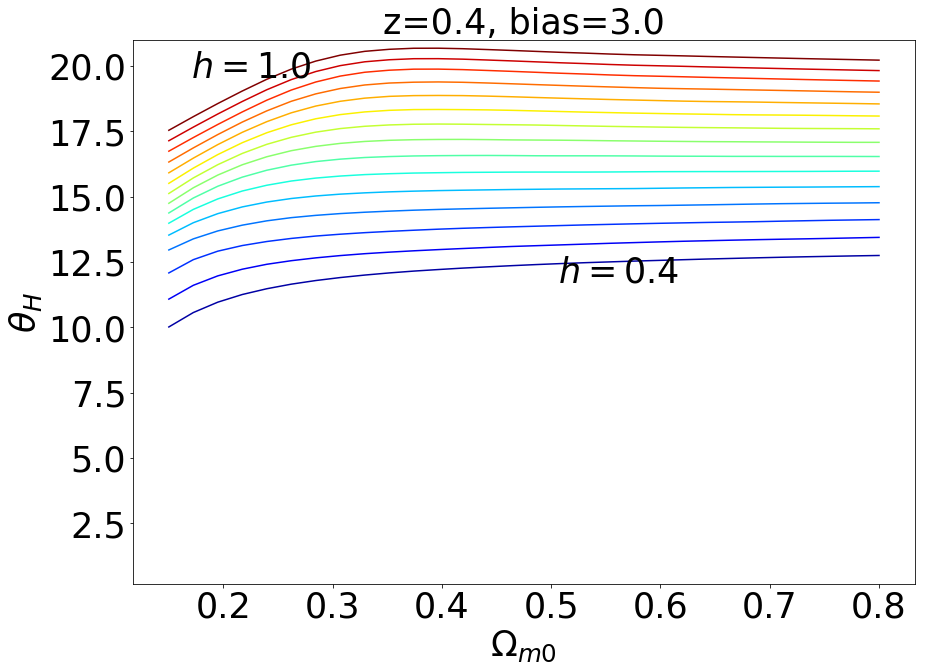}
	\includegraphics[width=0.33\textwidth]{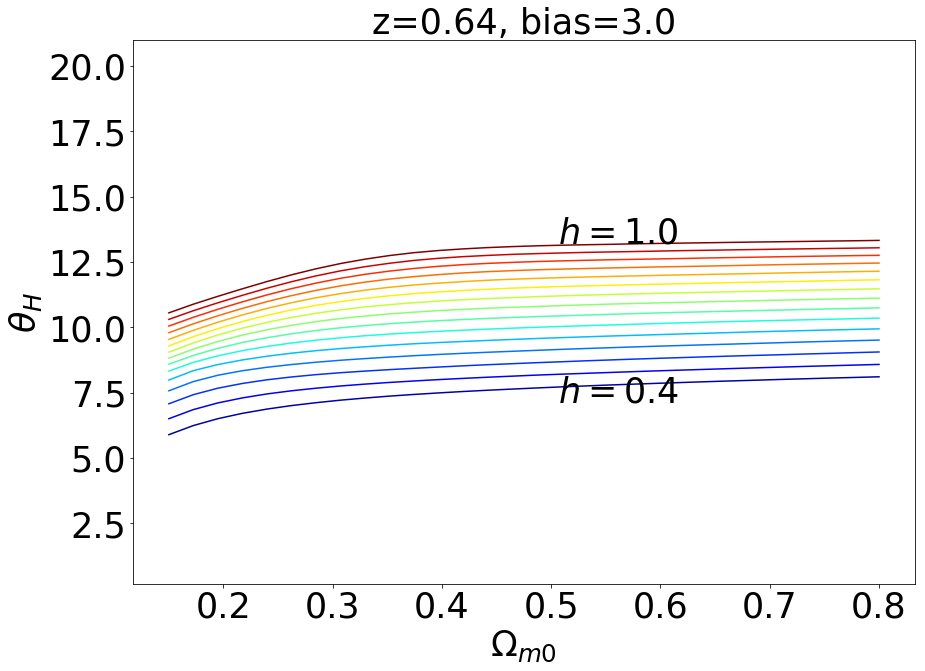}
 	\includegraphics[width=0.33\textwidth]{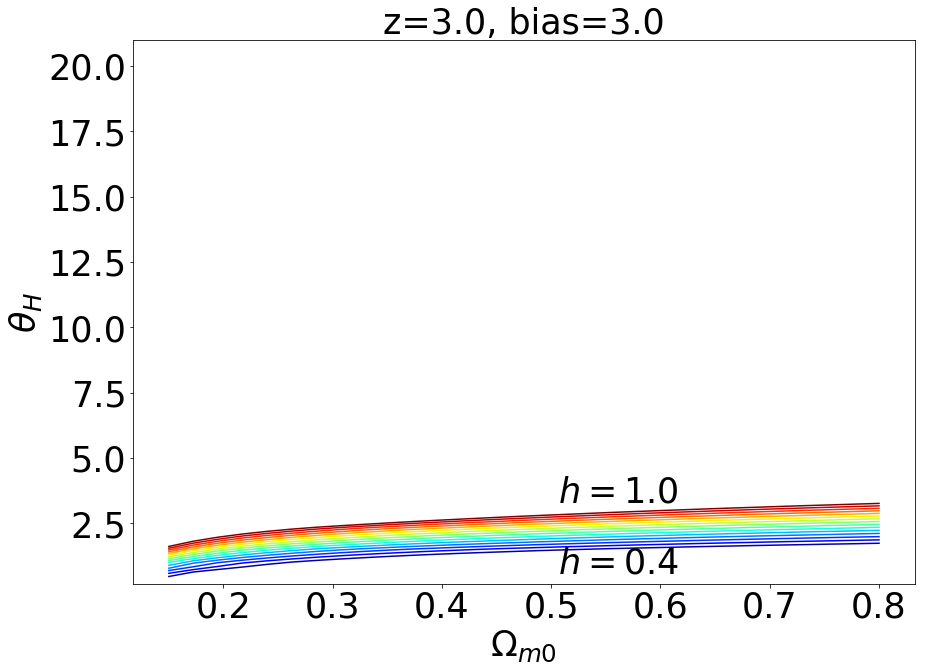}
	\caption{The angular homogeneity scale $\theta_H$ as a function of the present-day matter density parameter $\Omega_{m0}$ for the Hubble Constant $h \in [0.4,1.0]$ at $z=0.4$, $0.64$, $3.0$. We can see that $\theta_H$ is a monotonically increasing function for all values of $\Omega_{\rm m0}$ for $z \geq 0.64$.}
	\label{fig:0.4hom_2}
\end{figure*}  

\begin{figure*}[!ht]
	\centering
	\includegraphics[width=0.33\textwidth]{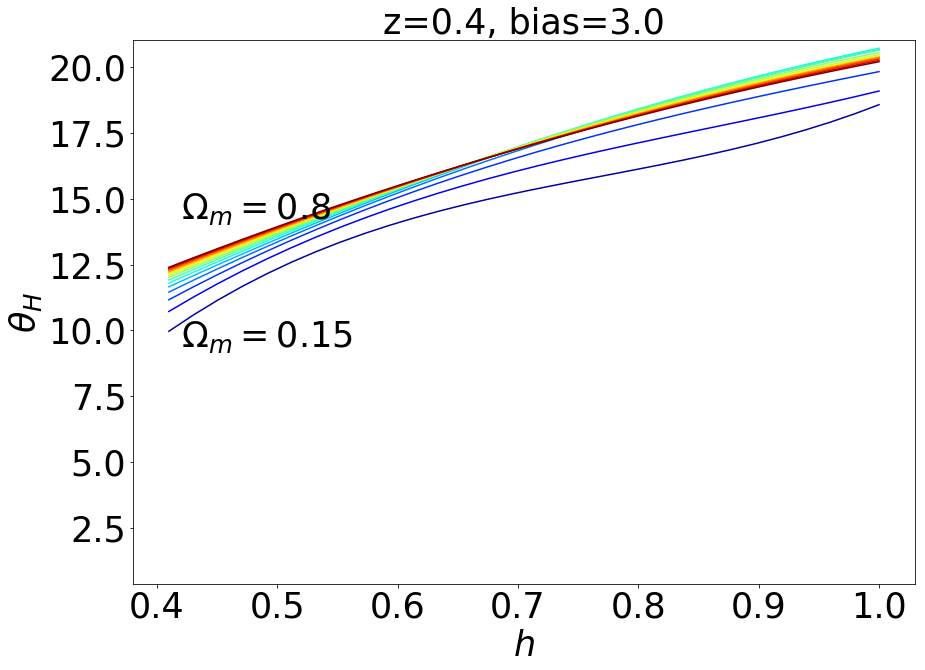}
	\includegraphics[width=0.33\textwidth]{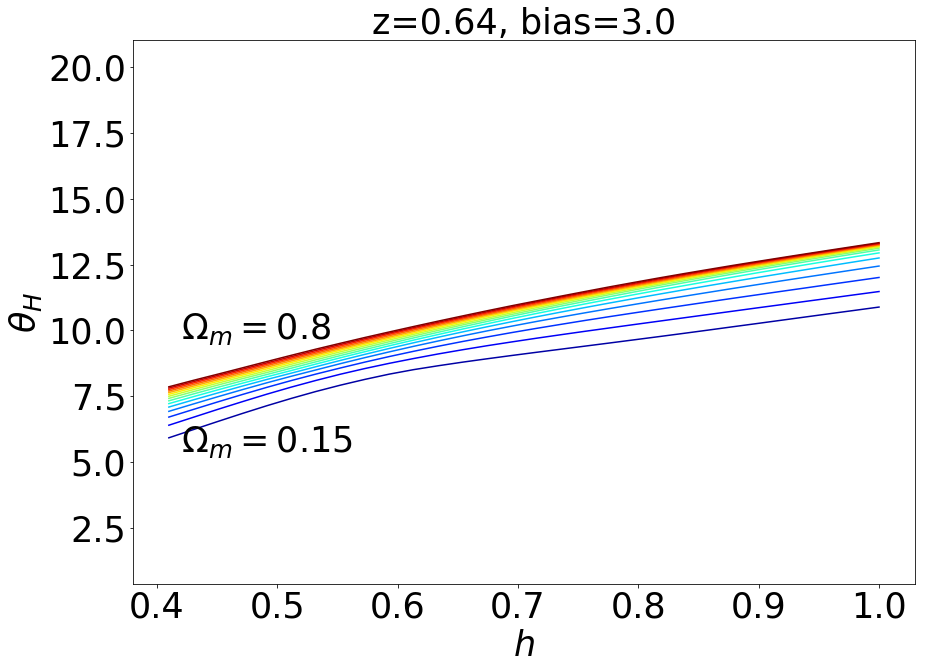}
 	\includegraphics[width=0.33\textwidth]{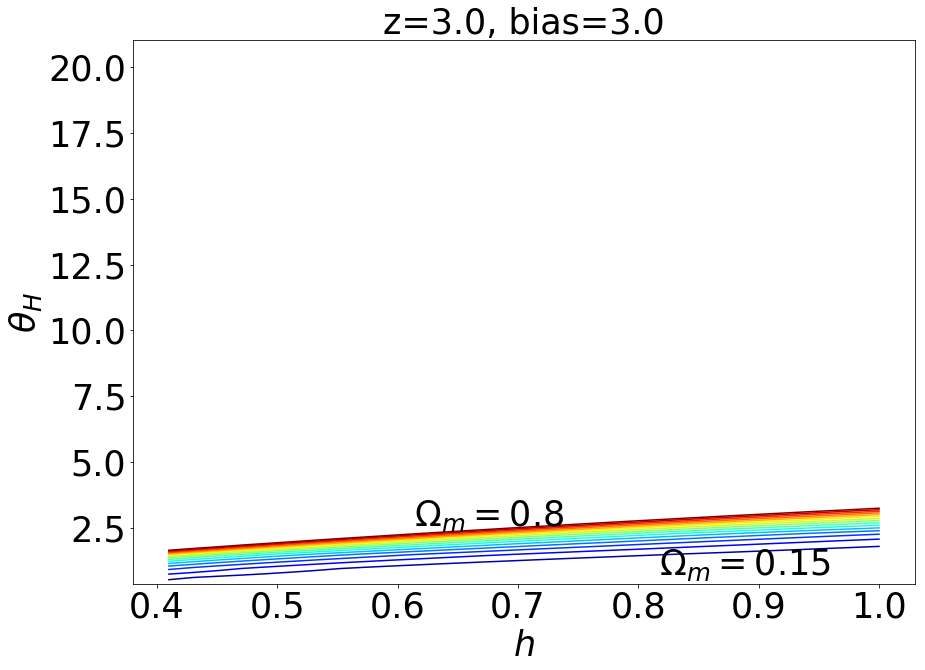}
	\caption{The angular homogeneity scale $\theta_H$ as a function of the Hubble Constant $h$ for $\Omega_{\rm m0} \in [0.15,0.8]$ at $z=0.4$, $0.64$, $3.0$. Clearly, $\theta_{\rm H}$ is a monotonically increasing function for all values of $h$, similarly to the $R_{\rm H}$ case.}
	\label{fig:0.4hom}
\end{figure*}

For a novel result we compute the theoretical predictions of the 2D angular homogeneity scales $\theta_{\rm H}$ by Eq.~\eqref{eq:d2xi}. Following the same token of the spatial homogeneity scale analysis, we display the 2D fractal dimension in the right panel of Fig.~\ref{fig:3d0.4ksi}, again for the redshift values $z=0.4$, $0.64$, and $3.0$, assuming $\Omega_{\rm m0}=0.31$ and $h=0.677$, where the dashed line corresponds to a deviation of $1\%$ from angular homogeneity, as in Eq.~\eqref{e1.98}.

We set the width of each redshift bin to $0.01$, which corresponds to the width of the spherical shell. The size of the bin is chosen to avoid projection effects, in order to presents minimal bias effects on the homogeneity scale measurements, as well as to increase the number of data points, thus providing a good statistical performance for the analysis \cite{Goncalves:2018sxa, Goncalves:2020erb, Alonso:2013boa, Andrade:2022imy}.

In Fig.~\ref{fig:0.4hom_2}, we present the angular homogeneity scale $\theta_{\rm H}$ as a function of $\Omega_{\rm m0}$, within the range $\Omega_{\rm m0}=[0.15,0.8]$, for redshifts $z=0.4$, $0.64$, $3.0$, and different values of $h$. Our results show that the function $\theta_H$ does not exhibit a one-to-one correspondence for the $\Omega_{m0}$ values within the specified interval and lower redshift. Nonetheless, we should remark that this behaviour changes for higher redshift ranges. For instance, for $z=0.64$ and $z=3.0$ our numerical results indicate that the angular homogeneity scale does exhibit a one-to-one correspondence with $\Omega_{\rm m0}$ ranging from $0.15$ to $0.8$, conversely from its 3D counterpart.

We also analyzed the angular homogeneity scale $\theta_{\rm H}$ as a function of $h$ (Fig.~\ref{fig:0.4hom}) within the range $h=[0.4,1.0]$, for redshifts $z=0.4$, $0.64$, $3.0$ and different values of $\Omega_{\rm m0}$. The results show that $\theta_{\rm H}$ does exhibit monotonicity concerning the range of $h$ assumed, keeping the same possibilities that it 3D counterpart. Also, both results, in Fig.~\ref{fig:0.4hom_2} and Fig.~\ref{fig:0.4hom}, show that $\theta_{\rm H}$ tightly depends on redshift.

Hence, our results demonstrate that $\theta_{\rm H}$ can be used as a cosmological test for the parameter $h$ and for $\Omega_{\rm m0}$ given that the redshift is high enough (e.g. $z = 0.64$), in order to break the non-monotonicity between $\Omega_{m0}$ and $\theta_H$. A similar behavior is found for the $z=3.0$ case. 

For the 2D homogeneity scale $\theta_{\rm H}$, we see a relative change of $\sim3.4\%$, $\sim5.6\%$ and $\sim17.9\%$ at redshift $z=0.4$, $0.64$, and $3.0$, respectively, when assuming reasonable values of $\Omega_{\rm m0}$, i.e., from $0.21$ to $0.41$. So, the $\theta_{\rm H}$ only significantly changes with respect to $\Omega_{\rm m0}$ at high redshifts. On the other hand, $\theta_{\rm H}$  significantly changes with respect to $h$ values at reasonable values, i.e., $h=0.6$ to $h=0.8$ at different redshifts. We see a relative change of $\sim 10.4\%$, $\sim 10.11\%$, and $\sim 13.9\%$ at the redshifts $z=0.4$, $0.64$, and $3.0$, respectively. Thus, we  show the 2D homogeneity scale $\theta_{\rm H}$ is more sensitive to the Hubble Constant than the 3D homogeneity scale.

We emphasize  that all results shown above were obtained by fixing the clustering bias to $b=3.0$. However, we performed analyses assuming other values for the bias parameter, and our conclusions were unchanged. In Fig.~\ref{fig:bias643} of Appendix, we present the angular homogeneity scale $\theta_{\rm H}$ as a function of the present-day matter density parameter $\Omega_{\rm m0}$ for bias factor $b \in [0.2,4.0]$ at $z=0.64,3.0$. We obtain that $\theta_{\rm H}$ is a monotonically increasing function for all values of $\Omega_{\rm m0} \in [0.15,0.8]$ for $z>0.6$, and this relationship is independent of the bias value. So, our analysis reveals that $\theta_{\rm H}$ {does} exhibit a one-to-one correspondence with the present-day matter density parameter $\Omega_{\rm m0}$ and the Hubble Constant for $z \gtrsim 0.6$.

\subsection{Current constraints from $\theta_{\rm H}(z)$}
\label{sec:mcmc}
$ $

In addition to the theoretical calculations presented by Eq.~\eqref{e1.98} and Eq.~\eqref{eq:d2xi}, we use observational data from redshift surveys to estimate $D_2(\theta)$ and $\hat{\omega}_{ls}$ by means of Eq.~\eqref{eq:defD2} and Eq.~\eqref{est}. Thus, we can measure $\theta_{\rm H}(z)$ from galaxies catalogs, without assuming a fiducial cosmology to convert redshifts into cosmological radial distances ($r$), as we only use the celestial coordinates of the catalogs.

Here, we directly use the three currently available measurements of $\theta_{\rm H}$ at the $z \geq 0.64$ reported in~\cite{Andrade:2022imy}, which is shown in Table~\ref{tab:t10}. These measurements were obtained from 107,500 luminous red galaxies within the redshift range $0.6 < z < 1.0$, with a sky coverage of $2,566 \; {\mathrm{deg^{2}}}$, from the Data Release 16 of the Extended Baryon Oscillation Spectroscopic Survey (eBOSS)~\cite{dr16}.

\begin{table}[]
\centering
    \caption{The redshift bins adopted in \cite{Andrade:2022imy}, along with the redshit bin means, the number of LRGs in each bin and angular homogeneity scale data points.}
    \label{tab:t10}
\begin{tabular}{ccccc}
\hline 
z & $\bar{z}$ & $N_{gal}$ & $\theta_{\rm H}$\;($\mathrm{deg}$) \\ 
\hline 
0.67 - 0.68 & 0.675 & 4209 & 7.57 $\pm$ 2.91 \\ 
0.70 - 0.71 & 0.705 & 4073 & 7.49 $\pm$ 2.63 \\ 
0.73 - 0.74 & 0.735 & 4210 & 8.88 $\pm$ 2.81 \\ 
\hline 
\end{tabular} 
\end{table}

\begin{figure}[!t]
	\centering
\includegraphics[width=0.49\textwidth]{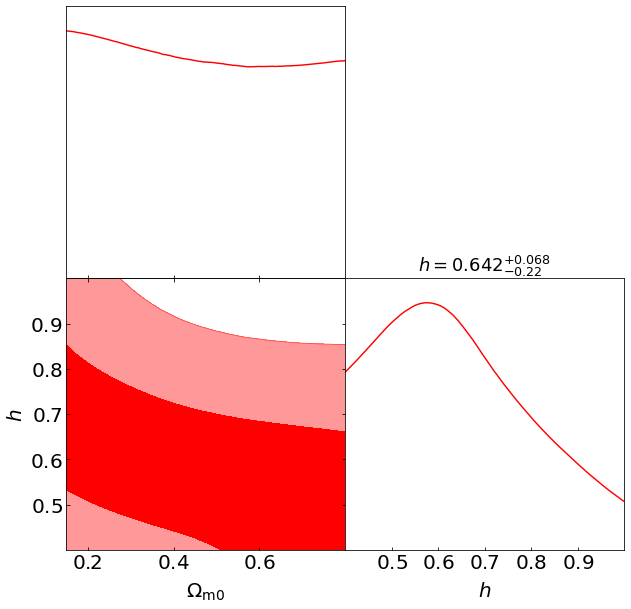}
	\caption{Constraints on the $\Omega_{\rm m0}-h$ plane from the three $\theta_{\rm H}(z)$ measurements reported in~\cite{Andrade:2022imy}.}
	\label{fig:3p4p2p_b}
\end{figure}

To illustrate its feasibility of constraining the $\Omega_{\rm m0}-h$ plane, we perform a Markov Chain Monte Carlo (MCMC) analysis considering the intervals $\Omega_{m0}\in [0.15, 0.8]$ and $h \in[0.4, 1.0]$. In this analysis, $\Omega_{\rm m0}$ and $h$ are the free cosmological parameters, and $b$ is the clustering bias factor for the galaxies survey, that is fixed in $b = 3.0$, for simplicity.

Given the small number of $\theta_{\rm H}(z)$ measurements  presently available, we focus on the $\Omega_{\rm m0}-h$ plane and the results are shown in Figure~\ref{fig:3p4p2p_b}.  At 1$\sigma$ confidence level, we find $h = 0.642^{+0.068}_{-0.220}$ which corresponds to a relative error of $\varepsilon_h = \sigma_h/h \sim 22\%$. However, $\Omega_{\rm m0}$ is poorly constrained, since we only have three data points at the $0.67<z<0.74$ range, and as shown before, $\theta_{\rm H}$ is not significantly sensitive to $\Omega_{\rm m0}$ at that range -- only at higher one, which we have no measurements at the moment. This is consistent with our analysis in subsection~\ref{sec:thetah_results} that $\theta_{\rm H}$ has a stronger constraining power on the Hubble Constant than $\Omega_{\rm m0}$.

Although the current $\theta_{\rm H}(z)$ constraints are not competitive when compared with current CMB limits (with $\varepsilon_h \sim 0.6\%$~\cite{aghanim2021planck}) and type Ia Supernovae (with $\varepsilon_h \sim 1.9\%$), they show a good constraining power over $h$, in agreement with the theoretical results presented in the previous sections, i.e, the homogeneity scale $\theta_{\rm H}$ being sensitive to $h$ values. 

As BAO data is sensitive to $\Omega_{\rm m0}$, a joint analysis with $\theta_{\rm H}$ measurements may be used to study a break of the parameter degeneracy of BAO for $H_0 r_{\rm drag}$. We will perform such study in future work, where we also plan to perform an analysis with an extended $\theta_{\rm H}$ measurement compilation, spanning a wider redshift range -- as we need to re-assess the $\theta_{\rm H}$ measurements available in the literature using the same angular homogeneity scale estimator ($D_{2}(\theta)$), and 2-point angular correlation function definition.

\section{Conclusions}
\label{sec:conclusions}

In this paper, we revisited, complemented and extended the discussion about the feasibility of measurements of the homogeneity scale as a cosmological test. Assuming a spatially flat $\Lambda$CDM model, we conducted a comparative analysis of the behavior of both spatial ($R_{\rm H}$) and angular ($\theta_{\rm H}$) homogeneity scales as a function of the redshift.

In agreement with \cite{Nesseris:2019mlr}, our first results showed that the spatial homogeneity scales do not exhibit a one-to-one behavior for reasonable values of $\Omega_{\rm m0}$, as estimated by current observations. However, our results showed a monotonic behavior of $R_{\rm H}$ for reasonable values of the Hubble constant within a large redshift interval. Actually, $R_{\rm H}$ was found to be influenced by $h$, with a variation of $\sim$ 3.7\%, $\sim$ 4.3\%, and $\sim$ 9.8\% at redshifts of 0.4, 0.64, and 3.0, respectively.

Concerning to the angular homogeneity scale, we showed that $\theta_{\rm H}$ can be used as a cosmological probe for $h$ at any redshift. Furthermore, for $z \gtrsim 0.6$, we also showed that the angular homogeneity scale is not only a one-to-one function for reasonable intervals of $\Omega_{\rm m0}$ and $h$, but it is also quite sensitive to values of $h$, varying by $\sim10.4\%$, $\sim10.11\%$, and $\sim13.9\%$ at redshifts 0.4, 0.64, and 3.0, respectively. To illustrate the feasibility of the $\theta_{\rm H}(z)$ test in constraining the parameters $\Omega_{\rm m0}$ and $h$, we performed an MCMC analysis considering the three currently available data points in the literature~\cite{Andrade:2022imy}. Our analysis found $h = 0.642^{+0.068}_{-0.22}$ at 1$\sigma$ confidence level. However, $\Omega_{\rm m0}$ is mostly unconstrained as $\theta_{\rm H}$ is not sensitive to this parameter.

Finally, it is important to mention that $\theta_{\rm H}$ is measured from cosmological observations in a model-independent way, differently from its 3D counterpart $R_{\rm H}$. Therefore, the results of this paper, showing the one-to-one behavior of $\theta_{\rm H}$ with $\Omega_{\rm m0}$ and $h$ for intermediate and high redshifts and the potential constraining power of the $\theta_{\rm H}(z)$ measurements, are crucial to ensure a new, model-independent way to estimate cosmological parameters. This kind of observable is particularly important given the current tensions of the standard cosmology and the many alternatives to solve them.

\section*{Acknowledgements}

XS acknowledges financial support through a PhD fellowship from the Coordena\c{c}\~ao de Aperfei\c{c}oamento de Pessoal de N\'{\i}vel Superior (CAPES). CB acknowledges financial support from Funda\c{c}\~ao Carlos Chagas Filho de Amparo \`a Pesquisa do Estado do Rio de Janeiro (FAPERJ) -  Postdoc Nota 10 fellowship. UA has been supported by Department of Energy under contract DE-FG02-95ER40899, and Leinweber Center for Theoretical Physics at the University of Michigan. JSA is supported by Conselho Nacional de Desenvolvimento Cientifico e Tecnol\'ogico (CNPq 307683/2022-2) and FAPERJ grant 259610 (2021).

\bibliographystyle{unsrt}
\bibliography{homogeneityscale}



\appendix



\section*{Appendix}
\label{sec:hg}

For completeness, we present below numerical results for the angular homogeneity scale $\theta_{\rm H}$ as a function of the present-day matter density parameter $\Omega_{\rm m0}$, fixing Hubble constant $h$ at $0.6766$ for bias $\in[0.2,4.0]$ at $z=0.64,3.0$.

\begin{figure*}[!t]
	\centering
	\includegraphics[width=0.49\textwidth]{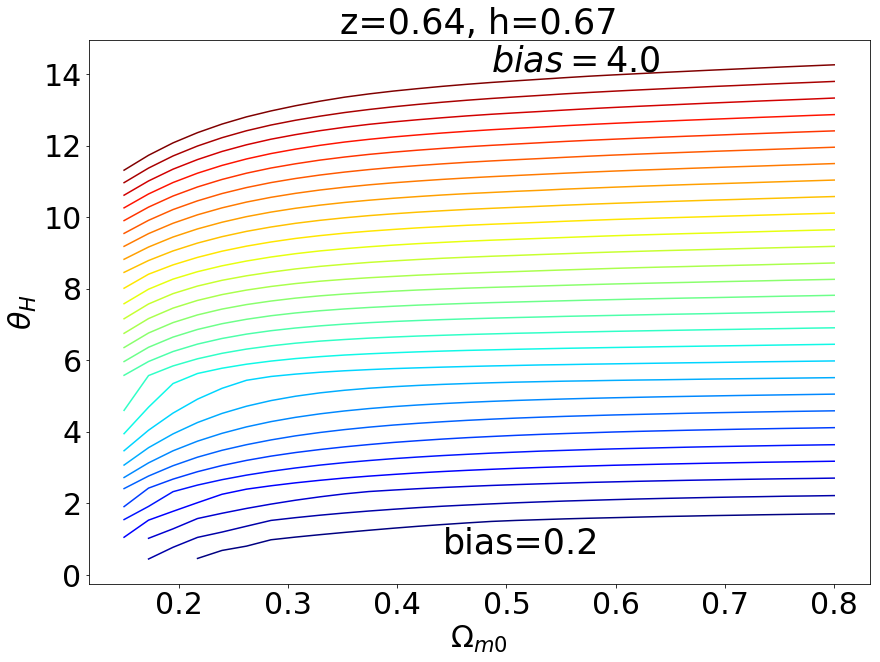}
    \includegraphics[width=0.49\textwidth]{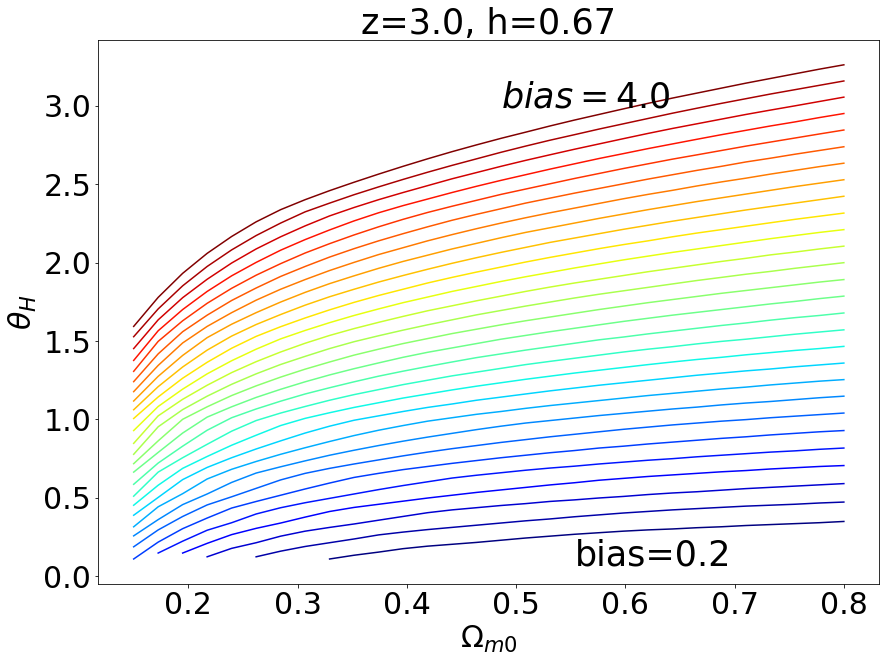}
	\caption{The angular homogeneity scale $\theta_{\rm H}$ as a function of the present-day matter density parameter $\Omega_{\rm m0}$ fixing hubble constant $h$ at 0.6766 for bias $\in[0.2,4.0]$ at $z=$0.64, 3.0 .
    $\theta_{\rm H}$ is a monotonically increasing function for all values of  $\Omega_{\rm m0}$ when redshift is larger than 0.64, which is independent of the bias factor.}
	\label{fig:bias643}
\end{figure*}

\end{document}